\documentclass[preprint, showpacs, aps,pre]{revtex4}
\usepackage{ulem}
\begin{document}

\title{Nanometric pitch in modulated structures of twist-bend nematic liquid crystals}

\author{M. P. Rosseto$^1$\email{Corresponding Author: michelyp.rosseto@hotmail.com}, R. R. Ribeiro de Almeida$^2$, R. S. Zola$^2$, G. Barbero$^{3,4}$, I. Lelidis$^5$, and L. R. Evangelista$^1$}
\affiliation{
$^1$ Departamento de F\'isica, Universidade Estadual de Maring\'a\\ Avenida Colombo, 5790-87020-900 Maring\'a, Paran\'a, Brazil.\\
$^2$ Universidade Tecnol\'ogica Federal do Paran\'a, Campus Apucarana,\\
Rua Marc\'ilio Dias 635, 86812-460 Apucarana, Paran\'a, Brazil.\\
$^3$ Department of Applied Science and Technology,
Politecnico di Torino, Corso Duca degli Abruzzi 24, 10129 Torino, Italy.\\
$^4$ National Research Nuclear University MEPhl (Moscow Engineering Physics Institute), Kashirskoye shosse 31, 115409 Moscow, Russian Federation.\\
$^5$ Solid State Section, Department of Physics, University of Athens, Panepistimiopolis, Zografos, Athens 157 84, Greece.
}

\begin{abstract}
The extended Frank elastic energy density is used to investigate the existence of a stable periodically modulate structure that appears  as a ground state exhibiting a twist-bend molecular arrangement. For an unbounded sample, we show that the  twist-bend nematic  phase $\rm N_{TB}$ is characterized by a heliconical structure with a pitch in the nano-metric range, in agreement with experimental results. For a sample of finite thickness, we show that the wave vector of the stable periodic structure depends not only on the elastic parameters but also on the anchoring energy, easy axis direction, and the thickness of the sample.
\end{abstract}
\pacs{61.30.Dk,61.30.Gd}
\maketitle

\section{Introduction}
\label{Sec:Introd}

Modulated materials are remarkable examples of how structure defines macroscopic properties in condensed matter. From naturally occurring phononic activity in crystals to synthetic superconductor superlattices~\cite{modulated}, modulated structures are key elements and fine tailored in cutting edge applications. They occur naturally in calamitic liquid crystals such as layered arrangements (in smectics), helical organization in chiral nematics or even defect lattice structures in blue 
phase materials~\cite{jacgil}. Such organization defines, for example, exciting photonic properties like the ones observed in chiral nematic and blue phases. These are not, however, the only possible modulated phases in liquid crystals. Newly synthesized materials open up possibilities for new structures quite often in the liquid crystals field. A new class of materials has been recently discovered in the liquid crystals (LC) field to show molecular organizations that permits one to characterize them as new orientational phases. The twist-bend nematic ($\rm N_{\rm TB}$)  is one of these phases. It has a double degenerate ground-state with a periodically modulated  heliconical structure exhibiting a nanoscale pitch favoured by achiral bent-shaped molecules. This phase has been experimentally observed  very recently in bent molecular 
dimers~\cite{Oleg,Adlem, Noel}, trimers~\cite{quan} and, more recently yet, in  rigid bent-core 
materials~\cite{Min} and in chiral dimers as well~\cite{chiraldop1}. In a very short time, such materials have already found applications, as reported in~\cite{jie,mh}. It has been theoretically predicted years ago~\cite{Meyer, Dozov, Memmer} together with a different orientational, modulated  structure designed as splay-bend phase~\cite{Dozov}.  From the theoretical point of view, besides the initial 
predictions~\cite{Dozov,Memmer}, it has been argued~\cite{selinger} that flexoelectricity could be the driving force for the formation of the modulated phase.  Nonetheless, recent studies indicate that the flexoelectric model cannot explain the large compression modulus of the $\rm N_{\rm TB}$ phase and that the driving force for the $\rm N_{\rm TB}$ phase comes from steric interactions~\cite{new2}.

One remarkable point regarding the $\rm N_{\rm TB}$ phase is the fact that it presents, in the ground state, spontaneous twist and bend deformations, exhibiting  an oblique helicoidal structure with a nanometric pitch, which  is directly  related  to a bend elastic constant $K_{33}$  much smaller than both $K_{22}$ (twist) and $K_{11}$ (splay). However,  such ratios of the elastic constants have not been detected experimentally  in the ground state region. To date, experimental investigations of elastic properties in the  $\rm N_{TB}$ phase have been reported by using  the extrapolation technique to  extract $K_{33}$, $K_{22}$ and $K_{11}$ elastic constants of those  materials in the region close to the phase transition. Thus, the  approach to explore these elastic constants is still based on  the standard elastic energy density,  obtaining the  threshold of the Fr\'{e}edericksz transitions  in splay, bend and twist geometries~\cite{freed}.

In this paper, we use a recently proposed elastic model~\cite{pre,lc} to investigate the possibility of a ground state of  periodic modulations exhibiting heliconical orientational ordering. For an infinite sample, we show that the pitch of the modulated structure may be in the nanometric range, as experimentally found. For a finite-length situation, when the  interaction between the surface and the twist deformation is  negligible, by means of the transversality condition, we demonstrate that the mechanical torque transmitted to the surface in equilibrium vanishes. In this case the wave vector of the periodic modulation coincides with the one found for the infinite sample. In the presence of an anchoring energy at the surface, we show that this wave vector depends on surface properties like easy axis angle and anchoring strength as well as on the bulk properties like the elastic parameters, cone angle, and the thickness of the sample, a behavior expected in heliconical structures,  as observed in 
cholesterics~\cite{jacgil,jump}. The paper is organized as follows: In 
Secs.~\ref{Sec:Elastic} and~\ref{Sec:Pitch}, the elastic model for the nematic twist-bend is revisited in order to show that the conditions for the stability of the phase supports the existence of a pitch in the nanometric range, as experimentally verified. In 
Sec.~\ref{Sec:Finite}, the analysis focuses on a nematic twist-bend phase confined to a sample of finite thickness to explore the surface effects on the wave vector of the modulated structure. Some concluding remarks are drawn in Sec.~\ref{Sec:Conclusion}.

\section{The elastic model}
\label{Sec:Elastic}

The general elastic free energy is characterized  by a director ${\bf n}$ and a helix axis ${\bf t}$ related to the chiral twisted collective arrangement~\cite{pre}. One assumes furthermore no polar order, in such a manner that ${\bf n}$ is the usual nematic director. As shown previously~\cite{Libro, pre}, the free energy density compatible with the symmetry of the phase, in absence of an external field,  may be written as

\begin{widetext}
\begin{eqnarray}
\label{fzao}
f &=& f_0 - \frac{1}{2} \eta ( {\bf n} \cdot {\bf t})^2  + \kappa_1 \,{\bf t}\cdot \left[{\bf n}\times(\nabla \times {\bf n})\right]+\kappa_2\,{\bf n}\cdot (\nabla \times {\bf n}) \nonumber + \kappa_3 ({\bf n}\cdot{\bf t}) (\nabla \cdot{\bf n})
 + \frac{1}{2}K_{11} (\nabla \cdot {\bf n})^2 \nonumber \\
&+& \frac{1}{2} K_{22} \left[{\bf n} \cdot (\nabla \times {\bf n})\right]^2
+ \frac{1}{2} K_{33} ({\bf n} \times \nabla \times {\bf n})^2
 - (K_{22} + K_{24}) \nabla \cdot ({\bf n} \nabla \cdot {\bf n} + {\bf n} \times \nabla \times {\bf n}) \nonumber \\
&+& \mu_1 [{\bf t} \cdot ({\bf n} \times \nabla \times {\bf n})]^2 + \nu_1 [{\bf t} \cdot \nabla ({\bf t} \cdot {\bf n})]^2
+ \nu_2 [{\bf t} \cdot \nabla ({\bf n} \cdot {\bf t}) (\nabla \cdot {\bf n})]
+ \nu_3 [\nabla ({\bf t}\cdot{\bf n})]^2 + \nu_4 [({\bf t} \cdot \nabla) {\bf n}]^2 \nonumber \\
&+& \nu_5 [\nabla ({\bf n} \cdot {\bf t})\cdot ({\bf t} \cdot \nabla) {\bf n}] + \nu_6 \nabla ({\bf n}\cdot{\bf t})\cdot (\nabla \times {\bf n}). \nonumber \\
\end{eqnarray}
\end{widetext}
In Eq.~(\ref{fzao}),  $\kappa_i$, for $i=1,2,3$ are the elastic parameters connected with the spontaneous splay, twist and bend, respectively, while $K_{11}$, $K_{22}$, $K_{33}$ and $K_{24}$ are the Frank elastic constants.
The new elastic parameters are $\eta$, which is a measure of the coupling strength between $\bf n$ and $\bf t$,
$\mu_1$ and $\nu_i$, for $i=1,2,..., 6$.

In the $\rm N_{\rm TB}$ phase, the  director $\mathbf{n}$ forms an oblique helicoid  around the direction
 $\mathbf{t}$, with a constant tilt angle $\theta_0$ fixed by molecular interaction forces. By taking ${\bf t} = {\bf u_z}$, the nematic director in this case is~\cite{Oleg}

\begin{equation}
{\bf n}= [\cos \phi (z) \,{\bf u_x}+ \sin \phi (z) \,{\bf u_y }] \sin \theta_0 + \cos \theta_0\, {\bf u_z}.
 \label{1a}
\end{equation}
As shown in Ref.~\cite{pre}, this director configuration simplifies the problem in such a way that only the following terms will contribute to energy density, namely:

\begin{eqnarray}
 &f_d= f_1 - \frac{1}{2}\eta ({\bf n} \cdot {\bf t})^2 + \frac{1}{2}K_{22}[{\bf n} \cdot (\nabla \times {\bf n)}+q_0]^2\nonumber \\
 &+ \frac{1}{2} K_{33} ({\bf n} \times \nabla \times {\bf n})^2+ \nu_4[{\bf t} \times (\nabla \times {\bf n})]^2,
 \label{a1}
\end{eqnarray}
where $f_1=f_0-(1/2)\, K_{22}q_0^2$ is a constant, and $\nu_4 [({\bf t}\cdot \nabla){\bf n}]^2 = \nu_4[{\bf t} \times (\nabla \times {\bf n})]^2$, because $({\bf t}\cdot \nabla){\bf n} = \nabla ({\bf n} \cdot {\bf t}) -
({\bf t}\times \nabla \times {\bf n})$, and ${\bf n} \cdot {\bf t} = \cos \theta_0$, which is constant.
Furthermore, in
Eq.~(\ref{a1}),  $q_0=\kappa_2/K_{22}$ is  the wave vector of the cholesteric phase \cite{jacgil}. The free energy density per unit area corresponding to~(\ref{1a}) becomes:
\begin{eqnarray}
\label{Giova}
f(\phi', x) &=& f_1 - \frac{1}{2}\eta (1 - x) + \frac{1}{2}K_{22} (\phi'x-q_0)^2 \nonumber \\
&-& \frac{1}{2} K_{33}\phi'^2(x^2-xb_0),\end{eqnarray}
in which $x=\sin^2 \theta_0$, $b_0=(1+2\nu_4/K_{33})$, and $\phi'= d\phi(z)/dz$. In a perfectly aligned nematic phase, $x=0$ and therefore no bend distortion occurs. In the cholesteric phase, $x=1$, and the medium will have no extra distortion only if $b_0=1$. On the other hand if $b_0 \ne 1$, for $x=1$, and $\phi=qz$, the free energy density~(\ref{Giova}) becomes

\begin{equation}
f = f_1 + \frac{1}{2} K_{22} (q-q_0)^2 + \nu_4 q^2,
\end{equation}
from which one derives that the wave vector,  of the stable periodic structure, is renormalized by the presence of
$\nu_4$, in such a way that

\begin{equation}
q= \frac{q_0 K_{22}}{K_{22} + 2 \nu_4}.
\end{equation}
This corresponds to a stable structure only if $K_{22} + 2\nu_4 >0$.
The energy density Eq.~(\ref{Giova}) describes  the cholesteric phase for $\nu_4=0$. Thus     $\lambda_{\rm C} = \pi/q_0$ is the cholesteric pitch which is usually  micrometric. It can be in the micrometric range even when $x\neq 1$,
 since  a conical state is allowed   for a  small $K_{33}$ elastic 
 constant~\cite{jie, Xiang}.

\section{Unlimited sample: Nanometric pitch}
\label{Sec:Pitch}

In order to analyze the stability of the spontaneous periodic deformation in the $\rm N_{\rm TB}$ phase, we consider first an unlimited sample.  By taking into account  that
there is a periodicity along the helix axis of the $\rm N_{\rm TB}$ phase, we may obtain  $\phi(z)$ that  minimizes the energy of one period~\cite{Priestley}, that means, the functional

\begin{equation}\label{FA}
\frac{F(\lambda)}{S}= \int_0^\lambda dz \left[\frac{\gamma}{2} - K_{22}q_0 x \phi'+ \frac{1}{2}\alpha \phi'^2 \right],
\end{equation}
where $S$ is the area of the system in the $(x,y)$ plane, $\gamma=2 f_1 - \eta (1 - x)+K_{22}q_0^2$, and
$\alpha= (K_{22}-K_{33}) x^2 + K_{33} x b_0$ is an effective elastic constant. In~(\ref{FA}), $\lambda$ is the wavelength of the periodic structure. The corresponding Euler-Lagrange equation permits to obtain a first integral in the form

\begin{equation}
-K_{22} q_0 x + \alpha \phi' = C,
\label{EL}
\end{equation}
where $C$ is an integration constant.  For simplicity, we
introduce the average energy density per period, $g$,  that, by taking into account Eqs.~(\ref{FA})
and~(\ref{EL}), after some algebra, assumes the following form

\begin{equation}
\centering
g(C)=\frac{F(\lambda)}{S\lambda} = \frac{\gamma}{2} - \frac{1}{2\alpha} (K_{22} q_0 x)^2 + \frac{1}{2} \frac{C^2}{\alpha}.
\end{equation}
Minimization of $g$ with respect to $C$  yields $C=0$.
This result indicates that the mechanical torque present in the sample vanishes identically, i.e., $\partial f/\partial \phi'=0$.  The solution is of the form $\phi(z) = q z$, where

\begin{equation}
\label{qg}
\phi'(z) = \frac{K_{22} q_0 x}{\alpha} = q_{\rm B}.
\end{equation}
In this case, Eq.~(\ref{Giova}) assumes the form:

\begin{eqnarray}
\label{Giovax}
f(q, x) &=& f_1 - \frac{1}{2}\eta (1 - x) + \frac{1}{2}K_{22} (q\, x-q_0)^2 \nonumber \\
&-& \frac{1}{2} K_{33}q^2(x^2-xb_0).
\end{eqnarray}
The stable configuration of the  phase may be investigated by minimizing~(\ref{Giovax}) in terms of the independent parameters $q$ and $x$ as $(\partial f/\partial q)_{q=q^*}=0$ and $(\partial f/\partial x)_{x=x^*}=0$, giving~\cite{pre}:

\begin{equation}
\label{qstar}
q^*=\pm\frac{\sqrt{\eta}}{\sqrt{b_0K_{33}}}
\end{equation}
for the wave vector, and

\begin{equation}
\label{xstar}
 x^*=-\frac{b_0K_{33}\mp K_{22}q_0\sqrt{b_0K_{33}/\eta}}{K_{22}-K_{33}}
\end{equation}
for the cone angle of $\bf n$ with $\bf t$. By combining~(\ref{qstar}) with~(\ref{xstar}), one easily shows that the equilibrium value $q^*=q_{\rm B}$ as determined by~(\ref{qg}). Thus, the stable periodicity coincides with the one obtained from the minimization of the free energy,  and the wavelength of the periodic structure, i.e., the pitch of the twist-bend phase, is simply given by

\begin{equation}
\lambda_{\rm B} =\frac{\pi}{q_{\rm B}}  = \lambda_{\rm C} \frac{\alpha}{K_{22} x}.
\end{equation}
 It may be rewritten by using the definition of $\alpha$ given above in order to obtain

\begin{equation}
\label{ratio}
\frac{\lambda_{\rm B}}{\lambda_{\rm C}} = x \left(1-\frac{K_{33}}{K_{22}}\right) + \frac{K_{33} + 2 \nu_4}{K_{22}}.
\end{equation}

An initial estimation of the order of magnitude of the pitch can be obtained as follows. Since $0 < \lambda_{\rm B}/\lambda_{\rm C} \le 1$ and $x << 1$ in the  $\rm N_{\rm TB}$ phase, one observes that the values of $\nu_4$ lie  approximately in the interval $ -K_{33}/2 < \nu_4 < (K_{22}-K_{33})/2$, i.e., they can be also negative for $K_{33} >0$.   The extrapolation of elastic constant gives an indication of the
value of the bend coefficient  in the twist-bend nematic,  $K_{33} \approx 0.5  \, $pN and $K_{22} \approx 6.5\,$pN, as reported by Adlem~\cite{Adlem}. Since $\theta_0 \approx 15^{\circ}$~\cite{Zhang}, one obtains $ x = \sin^2\theta_0 \approx 0.06$ and $K_{33}/K_{22} \approx 0.07$. If we take $\nu_4 \approx -K_{33}/4$, we obtain $\lambda_{\rm B}/\lambda_{\rm C} \approx 0.09$. This confirms that in the range in which $\nu_4$ is negative, even if small in absolute value, the pitch of the $\rm N_{\rm TB}$ phase may be nanometric,  as it is evidenced in experimental reports on $\rm N_{\rm TB}$ material.

\section{Finite-length situation}
\label{Sec:Finite}

The analysis will be now focused on the situation in which the sample is limited by two flat surfaces, parallel to the $x,y$-plane, placed at $z=0$ and $z=d$ of a Cartesian reference frame. We assume that the value of the tilt angle at the surface is fixed and is equal to the one imposed by intermolecular forces. In the case in which the anchoring is strong on the lower surface, i.e., $\phi(0) =0$, and the upper surface imposes an easy axis $\Phi$, with a finite azimuthal anchoring energy, $W$, the  total energy of the sample is

\begin{equation}
\label{Giovaxx}
F = \int_0^d f(\phi', x) \, dz + \frac{1}{2} W \left[\phi(d) - \Phi \right]^2.
\end{equation}
In Eq.~(\ref{Giovaxx}), we have approximated the surface energy with a parabolic potential. Note that it is not a $\pi -$periodic potential.
The Euler-Lagrange equation for this system is (see Eq.~(\ref{qg}))
Suppose that the zenithal anchoring energy is zero but the  anchoring is finite.
\begin{equation}
\label{ELL}
\alpha \phi''(z) = 0,
\end{equation}
whose solution is again $\phi(z) = q z$ and, at the upper surface, it is subjected to the
boundary condition~\cite{Libro}:

\begin{eqnarray}
\left. \frac{\partial f}{\partial \phi'(z) } \right|_{z=d} &=& K_{22} x [\phi'(d) x - q_0] - K_{33} \phi'(d) (x^2 - xb_0)
\nonumber \\
&=& W [\phi(d) - \Phi].
\end{eqnarray}
When no interaction between the surface and the twist deformation is present,  $W=0$, the transversality condition
$\partial f/\partial \phi'  |_{z=d} =0$, i.e., the mechanical torque transmitted to the surface in the equilibrium state vanishes. Thus, the wave vector of the periodic modulation coincides with the one found for the unlimited system: $q=q_{\rm B}$.

We consider now the more realistic case of a bounded sample characterized by two easy axes in the presence of a finite anchoring energy. Let $\varphi(z) = \phi(z) - \Phi$, such that $\varphi(0) = - \Phi$ and $\varphi(d) = \phi(d)  - \Phi$. The free energy density~(\ref{Giovaxx}) becomes

\begin{eqnarray}
\label{Giovaxy}
F = f_0 d + \frac{1}{2}\alpha \int_0^d \left[\varphi'(z)^2 - 2 q_{B} \varphi'(z)\right]\, dz
+ \frac{1}{2}W \varphi(d)^2. \nonumber \\
\end{eqnarray}
The Euler-Lagrange equation has a first integral in the form (see again Eq.~(\ref{qg}))

\begin{equation}
\label{1star}
\varphi'(z) = q_{\rm B} + \delta q = q.
\end{equation}
The solution becomes $\varphi(z) = -\Phi + q z$, such that $\varphi(d) = q d - \Phi$. Substitution of (\ref{1star}) into (\ref{Giovaxy}) yields the simple expression

\begin{equation}
\label{fq}
F = f_0 d + \frac{1}{2} \alpha (q^2 - 2 q_{\rm B} q) d + \frac{1}{2}W (qd - \Phi)^2,
\end{equation}
which, after being minimized in $q$, yields:

\begin{equation}
\label{qq}
 q = \frac{q_{\rm B} + \left(W/\alpha\right) \, \Phi}{1 + \left(W/\alpha\right)\, d}.
\end{equation}
This is the wave vector of the stable periodicity for a finite-length sample when the upper surface is characterized by a weak anchoring  energy, $W$, and imposes an easy axis,  $\Phi$.  As it follows from Eq.~(\ref{FA}),  $\alpha$ plays the role of an effective elastic constant for the considered distortion. Hence $\ell=\alpha/W$ is the usual extrapolation length. From Eq.~(\ref{qq}), in the limit of zero anchoring energy ($d<<\ell$),  one retrieves the unlimited sample situation. That means the wavevector of the TB-phase remains equal to $q_B$.  To go on further, we  assume the following general expressions for the thickness of the sample and the easy axis, respectively:

\begin{equation}
\label{phiw}
d = m \lambda_{\rm B} + r \lambda_{\rm B} \quad {\rm and} \quad \Phi = m \pi + s \pi, \quad m = 0, 1, 2, \cdots,
\end{equation}
with $0 \le r \le 1$ and $0 \le s \le 1$.  From Eq.~(\ref{qq}),  we get for the relative variation of the actual pitch $q$ with respect to that in the infinite sample $q_B$ the expression

\begin{equation}
\label{qf}
\frac{q-q_B}{q_B}=\frac{(s-r)\pi}{q_B\,\ell+(m+r)\pi}\;.
\end{equation}
The limit of an infinite sample is achieved for $d\to \infty$, i.e., $m\to \infty$.   In this case, from Eq.~(\ref{qf}) we deduce that $q\to q_B$, as expected. The same limit is reobtained for $W\to 0$, i.e. $\ell\to \infty$. For a finite-length sample in the limit of strong anchoring ($W \to \infty$, and hence $\ell\to 0$), we  obtain

\begin{equation}
\frac{q-q_B}{q_B}=\frac{s-r}{m+r}.
\end{equation}
However, since for a macroscopic sample $m\gg 1$, then $q\sim q_B$. The conclusion is that for macroscopic sample the influence of the surface treatment on the actual pitch of the $N_{\rm TB}$ phase is small.

For infinite anchoring strength on both surfaces 
and parallel easy axes on them, if the thickness of the cell is not a semi-integer multiple of the pitch, one expects that the system will adapt by changing its natural wavevector to $q=q_B+\delta q$. That means, the pitch variation $\delta\lambda=\lambda_B/2N$, where $N=integer [d/\lambda_B]$. If one considers an ultrathin film of some dozens of pitch thick, the expected pitch variation is of the order of a few percent of $\lambda_B$.  When the anchoring is sufficiently weak at least at one surface, in case that the thickness of the cell is not a semi-integer multiple of  $\lambda_B$, the system can accommodate an incomplete turn, that is, the wave vector could remain equal to $q_B$.

The above analysis is similar to the problem for a cholesteric LC in a finite thickness cell with a preferred easy direction on the boundaries. In the latter case, the wavevector dependence on the anchoring and thickness of the cell, cholesteric pitch transition, and multistability have been extensively investigated~\cite{tim,anca,antonio} under the presence of an external field. Of course our present analysis, in the case of a finite thickness TB phase, is elementary and it intends just to find an upper limit for the variation of the free wavevector $q_B$.

\section{Concluding remarks}
\label{Sec:Conclusion}

We have  discussed twist-bend molecular organization by obtaining a link  between  the pitch of the   $\rm N_{\rm TB}$  phase and  the pitch of the cholesteric phase.  An  estimation of the order of magnitude of this pitch was done by  using the extrapolated values of the elastic parameters of the nematic twist-bend known in the literature. It shows that the elastic description we are using here predicts a stable ground state for heliconical twist-bend structure characterized by a nanometric pitch, as experimentally found. In addition, we have analyzed how the wave vector of the modulated phase depends on the thickness of the sample and also on the anchoring energy of the bounding surfaces. It is  shown that it is, actually, independent of the surface treatment if the sample is macroscopic. The analysis is a further indication that this elastic model may stand as one of the  pathways to enrich our understanding  of spontaneous  modulation in nematic liquid crystals systems.

\acknowledgments
This work has been partially supported by the National Institute of Science and Technology of Complex Fluids (INCT-Fcx, S\~ao Paulo, Brazil), and by the Competitiveness Program of NRNU MEPhl.


\begin{thebibliography}{99}

\bibitem{modulated} Ed. T. Tsakalakos, \textit{Modulated Structure Materials}, (Martinus Nijhoff Publishers, Dordrecht, 1984).

\bibitem{jacgil} P.G. de Gennes, J. Prost, {\it The Physics of Liquid Crystals}
(Clarendon Press, Oxford, 1993).

\bibitem{Oleg}  V. Borshch, Y-K. Kim,  J. Xiang,  M. Gao, A. Jakli, V.P. Panov, J.K Vij, C.T. Imrie, M.G. Imrie, G.H. Mehl and  O.D. Lavrentovich,   Nat.\ Commun.\ \textbf{4}, 2635 (2013).
\bibitem{Xiang} J. Xiang, S.V. Shiyanovskii, C. Imrie, and O.D. Lavrentovich
Phys. \ Rev. \ Lett. \textbf{112}, 217801 (2014).

\bibitem{Noel} D. Chen, J.H.  Porada, J.B. Hooper, A. Klittnick, Y. Shen, M.R. Tuchband, E. Korblova, D. Bedrov, D.M. Walba, M.A. Glaser, J.E. Maclennan and  N.A. Clark,  Proc.\ Natl.\ Acad.\ Sci.\ USA \textbf{110}, 15931 (2013).

\bibitem{Adlem} K. Adlem, M. Copic, G. R. Luckhurst, A. Mertelj, O. Parri, R.M. Richardson, B.D. Snow, B.A. Timimi, R.P. Tuffin, and D. Wilkes, Phys.\ Rev.\ E \textbf{88}, 022503 (2013).

\bibitem{quan} Y. Wang, G. Singh, D. M. A.-Kooijman, M. Gao, H. K. Bisoyi, C. Xue, M. R. Fisch, S. Kumarb and Q. Li, CrystEngComm \textbf{17}, 2778 (2015).

\bibitem{Min} D. Chen, M. Nakata, R. Shao, M. R. Tuchband, M. Shuai, U. Baumeister, W. Weissflog, D. Walba, M. A. Glaser, J. E. Maclennan, and N. A. Clark,  Phys.\ Rev.\ E \textbf{89}, 022506 (2014).

\bibitem{chiraldop1} C. T. Archbold, E. J. Davis, R. J. Mandle, S. J. Cowling and J. W. Goodby, Soft Mater \textbf{11}, 7547-7557 (2015).

\bibitem{jie} J. Xiang, Y. Li, Q. Li, D.A. Paterson, J. Storey, C.T. Imrie and O.D. Lavrentovich, Advanced Materials 27 (19), 3014 (2015).

\bibitem{mh} Y. Wang, Z.-G. Zheng, H. K. Bisoyi, K. G. G.-Cuevas, L. Wang, R. S. Zola and Q. Li, Materials Horizons, 3, 442 (2016).


\bibitem{Meyer} R. B. Meyer,  Les Houches Lectures, 1973, edited by R. Balian and G. Weill (Gordon and Breach, Les Houches 1976).

\bibitem{Dozov} I. Dozov, Europhys.\ Lett.\ \textbf{56}, 247 (2001).

\bibitem{Memmer}  R. Memmer,  Liq.\ Cryst.\ \textbf{29}, 483 (2002).

\bibitem{selinger} S. M. Shamid, S. Dhakal, and J. V. Selinger, Phys.\ Rev.\ E \textbf{87}, 052503 (2013).

\bibitem{new2} N. Vaupotic, S. Curk, M. A. Osipov, M. Cepic, H. Takezoe, and Ewa Gorecka, Phys.\ Rev.\ E \textbf{93}, 022704 (2016).


\bibitem{freed} V. Freedericksz, V. Tsvetkov, Phys.\ Z.\ Sov.\ Union \textbf{6}, 490 (1934).
\bibitem{pre} G. Barbero, L. R. Evangelista, M. Rosseto, R. S. Zola, and I. Lelidis, Phys.\ Rev.\ E \textbf{92}, 030501 (2015).

\bibitem{lc} R. S. Zola, G. Barbero,I. Lelidis, M. P. Rosseto and L. R. Evangelista, Liq Cryst.
2016;1–7. DOI:10.1080/02678292.2016.1229054.

\bibitem{jump} T. N. Orlova, R. I. Iegorov, and A. D. Kiselev, Phys.\ Rev.\ E. \textbf{89}, 012503 (2014).

\bibitem{Libro} G. Barbero and L. R. Evangelista, {\it An Elementary Course on the Continuum Theory for Nematic Liquid Crystals}, (World Scientific,  Singapore, 2001).

\bibitem{Priestley} P. Sheng, {\it Introduction to the Elastic Continuum Theory of Liquid Crystals } in {\it Introduction to Liquid Crystals}, Edited by E. B. Priestley, P. J. Wojtowicz, and Ping Sheng (Plenum Press, New York and London, 1979).

\bibitem{Zhang} R. R. Ribeiro de Almeida, C. Zhang, O. Parri, S. N. Sprunt and A. Jakli,  \textit{Liquid Crystals}, 2014, \textbf{41}, 1661.

\bibitem{tim} A.D. Kiselev, T.J. Sluckin, Phys.\ Rev.\ E. \textbf{71},031704 (2005).

\bibitem{anca} I. Lelidis, G. Barbero, A.L. Alexe-Ionescu, Phys.\ Rev.\ E. \textbf{87}, 022503 (2013).

\bibitem{antonio} A.M. Scarfone, I. Lelidis, G. Barbero, Phys.\ Rev.\ E. \textbf{84}, 021708 (2011).
\end{thebibliography}
\end{document}